Nanostructured complex oxides as a route towards thermal behavior in artificial spin ice systems

R.V. Chopdekar[1,*], B. Li[1], T.A. Wynn[1], M.S. Lee[1], Y. Jia[1], Z.Q. Liu[2,+], M.D. Biegalski[2], S.T. Retterer[2], A. T. Young[3], A. Scholl[3], and Y. Takamura[1]

[1]Department of Materials Science and Engineering, Univ. of California, Davis, Davis, California, 95616, USA

[2]Center for Nanophase Materials Sciences, Oak Ridge National Laboratory, Oak Ridge, Tennessee, 37831, USA

[3]Advanced Light Source, Lawrence Berkeley National Laboratory, Berkeley, California, 94720, USA

**Abstract**

We have used soft x-ray photoemission electron microscopy to image the magnetization of single domain $La_{0.7}Sr_{0.3}MnO_3$ nano-islands arranged in geometrically frustrated configurations such as square ice and kagome ice geometries. Upon thermal randomization, ensembles of nano-islands with strong inter-island magnetic coupling relax towards low-energy configurations. Statistical analysis shows that the likelihood of ensembles falling into low-energy configurations depends strongly on the annealing temperature. Annealing to just below the Curie temperature of the ferromagnetic film ($T_C$ = 338 K) allows for a much greater probability of achieving low energy configurations as compared to annealing above the Curie temperature. At this thermally active temperature of 325 K, the ensemble of ferromagnetic nano-islands explore their energy landscape over time and eventually transition to lower energy states as compared to the frozen-in configurations obtained upon cooling from above the Curie temperature. Thus, this materials system allows for a facile method to systematically study thermal evolution of artificial spin ice arrays of nano-islands at temperatures modestly above room temperature.

# I. Introduction

Frustration in magnetic systems occurs due to the inability to find a configuration that simultaneously minimizes all neighboring interactions, and can be found in systems with structural disorder (e.g. spin glasses) or systems with structural order but multiple competing interactions due to the geometry of the ordered state (e.g. pyrochlore structure crystals).[1,2] There has been considerable interest in the fabrication of two-dimensional analogues of a geometrically frustrated systems via an ensemble of nano-magnets to explore the physics of frustrated systems, including the effect of long-range dipolar interactions and the emergence of monopole-like defects.[3-7] The magnetostatic energy barrier for nano-magnet reversal can be considerably higher than room temperature, but effective thermodynamics of such athermal systems have been explored by enumerating the statistical likelihood of vertices with different magnetostatic energies.[8,9] If the nano-islands are thermally active near room temperature, thermally driven phenomena such as magnetic relaxation processes and monopole-like defect propagation can be measured with real-space imaging techniques. Prior magnetic microscopy experiments aimed at examining these effects in metallic nanostructures have achieved success in observing thermally-induced low energy magnetic configurations frozen-in during sample preparation[10]. Subsequent experiments have shown that ultrathin permalloy nano-islands have a thickness-dependent blocking temperature such that 3 nm thick islands yield a superparamagnetic blocking temperature near 300 K. [11] Artificial spin ice structures fabricated from ultrathin Fe layers embedded in Pd have also exhibited thermalization behavior at cryogenic temperatures.[12,13]

Experiments with ultrathin metallic nano-islands face challenges due to differences in blocking temperature between nominally identical islands caused by small changes in material thickness, as well as the possibility of rapid oxidation which can lead to suppression of magnetization on the timescales of experimental measurements. These issues can make it difficult to separate thermodynamic evolution of an artificial spin ice system from magnetization configuration changes due to island-island variation or materials degradation. In lieu of using nearly superparamagnetic islands, thermalization of artificial spin ice arrays in thicker permalloy islands was clearly demonstrated,[14] but these studies required temperatures above 800 K which introduced possible issues of interdiffusion of the magnetic material with the substrate as well as lateral diffusion of the nano-island material. Recent studies have shown that $Gd_{0.3}Co_{0.7}$ and $FePd_3$ (bulk Curie temperatures near 500 K) can be used in artificial spin ice geometries to exhibit thermally active behavior after appropriate annealing protocols.[15,16] In contrast, the use of a nanostructured epitaxial complex oxide such as $La_{0.7}Sr_{0.3}MnO_3$ (LSMO) could permit the study of thermally active artificial spin ice geometries just above room temperature due to a bulk Curie temperature of 370 K. Furthermore, LSMO exhibits good chemical stability in both vacuum and standard atmosphere environments up to 873 K,[17] permitting a broad range of annealing conditions. We explore thermal demagnetization in ensembles of coupled LSMO nano-islands and show that this artificial spin ice materials system offers a facile route to explore thermal effects in geometrically frustrated artificial systems at temperatures only slightly above room temperature.

**II. Experimental Methods**

Epitaxial LSMO films with 40 nm thickness were grown on conducting (001)-oriented Nb:SrTiO$_3$ substrates via pulsed laser deposition. Substrates were held at a deposition temperature of 700 °C in an ambient oxygen pressure of 200-300 mTorr with a laser fluence of 1 J/cm$^2$, and a post-deposition anneal in 300 Torr of oxygen minimized the presence of oxygen vacancies. The crystallinity of the films was characterized by high-resolution x-ray diffraction, with reciprocal space maps of the 103 reflection indicating that the films were wholly pseudomorphic. Following the LSMO deposition, ensembles of coupled nano-magnets were defined with a 70 nm thick chromium hard mask patterned via an electron beam lithography liftoff technique. A flood implant of 10$^{15}$ Ar$^+$ ions/cm$^2$ at an energy of 50 keV causes the regions of LSMO not protected by the Cr hard mask to be rendered amorphous and paramagnetic, while the crystalline LSMO protected by the mask retained its ferromagnetic order at room temperature.[18,19] At this dose, the magnetization of the implanted regions is reduced to less than 5% of the as-deposited crystalline LSMO as measured by SQUID magnetometry at 10 K. The Cr mask was removed from the sample after the implant and before subsequent measurements. Magnetoresistance measurements of a 500 nm wide and 20 micron long LSMO wire patterned using this technique (see supplemental Figure S1) gave a metal-insulator transition temperature of 338 K. This transition temperature is coincident with the Curie temperature due to the double-exchange mechanism in LSMO.[20]

The nano-island shape was chosen to be a stadium geometry (rectangle with semicircles at opposite ends), with total island length fixed at 470 nm while the widths of the islands were varied between 100 and 225 nm. The width variation served to control both the island shape anisotropy and the total magnetic moment, *m,* per island, but for all widths the magnetization was Ising-like and aligned along the island's long axis. While micron-scale (001)-oriented LSMO nanostructures have non-negligible magnetocrystalline anisotropy at low temperatures,[21] the magnetocrystalline anisotropy constant K$_1$ linearly decreases with increasing temperature and is close to zero near the Curie temperature.[22] Thus, the nanoscale islands in this work were found to be dominated by shape anisotropy at all measurement temperatures, and experimentally were single domain with the magnetization directed along the island long axis regardless of crystallographic orientation. The lattice spacing between neighboring kagome or square ice unit cells, *a,* was varied between 500 and 600 nm to tune the inter-island coupling strength $m^2/a^3$, which reflects the scale of the dipolar interaction.[23] This quantity can be related to the material's magnetization, $M(T)$, and the island volume, $V_{island}$, as $\frac{m^2}{a^3} = M^2(T)\frac{V_{island}^2}{a^3}$. Thus, for a fixed island volume and spacing, the coupling strength may also be tuned as a function of temperature. At room temperature, the inter-island coupling strength can be calculated based on a room temperature saturation magnetization of 225 emu/cm$^3$ for a continuous LSMO film of similar thickness. Room temperature coupling strengths based on this saturation magnetization and the above island spacings are quantitatively smaller to those obtained for thin permalloy and cobalt based spin ice systems,[23] and approach coupling strengths of thermally active ultrathin permalloy islands,[11] showing that the lithographically defined LSMO nano-island ensembles can also offer insights into the physics underlying thermal interactions in artificial spin ice systems.

Imaging of the magnetization orientation as a function of applied magnetic field and variable temperature was measured with Mn *L*-edge soft x-ray photoemission electron microscopy (X-PEEM) at the Advanced Light Source.[24] While magnetic contrast was measureable at all temperatures below room temperature, magnetization images were taken at a sample temperature of 110 K to minimize any thermal evolution of the magnetization and to maximize the x-ray magnetic circular dichroism (XMCD) contrast. A coil built into the sample holder allowed magnetic field pulses of up to 15 mT to be applied parallel to the in-plane projection of the x-ray incidence direction, with the nano-island magnetization imaged in remanence after any field pulse sequence. The x-rays impinged upon the sample at a 30° grazing angle with respect to the surface, and was set to be within 5° of the in-plane [100] crystal direction. The sensitivity axis of the X-PEEM imaging is parallel to the incoming x-ray direction, and is oriented horizontally from the right in the figures below. The white and black contrast indicates magnetization projections pointing to the right and left respectively, while a gray contrast indicates zero magnetization or a magnetization orthogonal to the sensitivity axis.[25] The nano-island magnetization direction, confined to be along the long axis of the island due to shape anisotropy, was determined from the projection of the XMCD contrast along the horizontal axis. Magnetic force microscopy at room temperature in atmosphere (supplemental information Figure S2) shows domain configurations consistent with those found with X-PEEM imaging, and no evidence of induced magnetization in the implanted matrix between neighboring nano-islands.

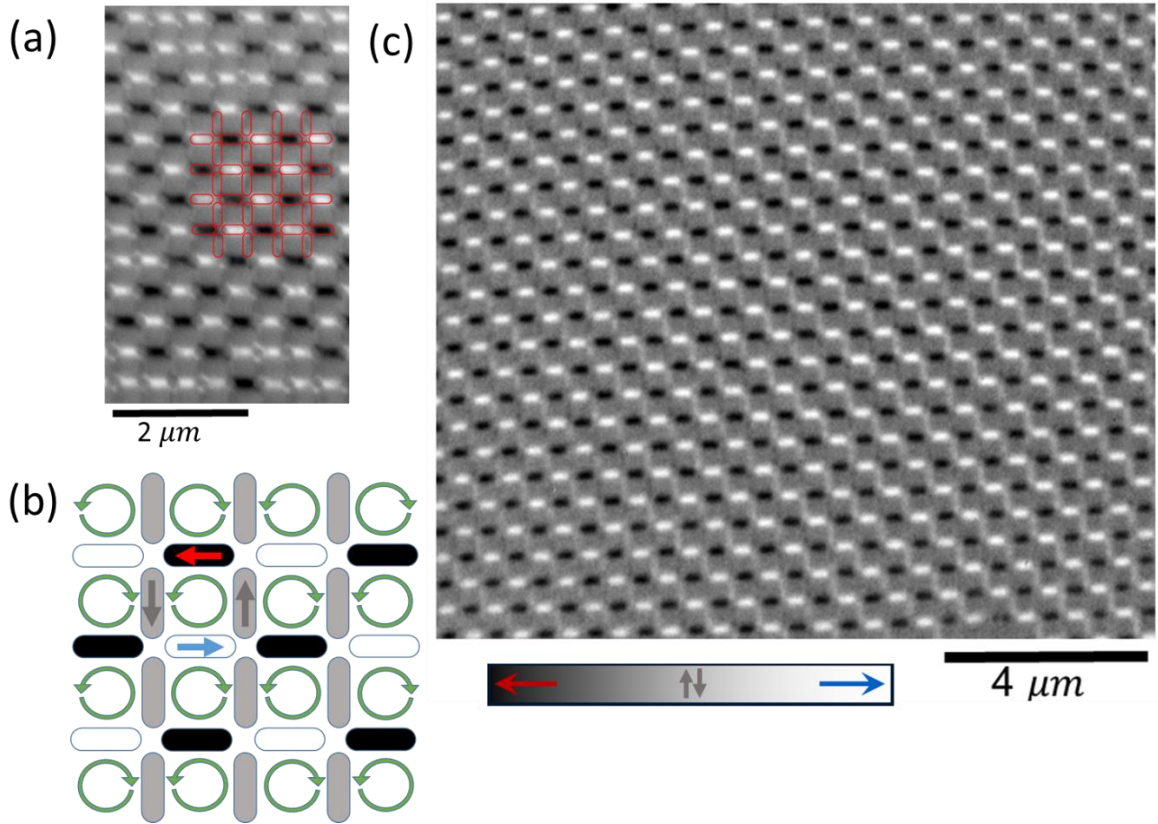

**Figure 1** (single column): (a) X-PEEM XMCD image of an artificial square ice array brought to ground state through a magnetic field demagnetization protocol. Epitaxial 470 nm x 175 nm x 40 nm LSMO nano-island locations are outlined in red. (b) Schematic of ground state nano-island magnetization configuration as seen with X-PEEM magnetic sensitivity axis along the horizontal direction parallel to the [100] crystallographic direction, with the red, gray and blue arrows indicating orientation of the magnetization of the nano-islands. (c) X-PEEM XMCD image of the entire artificial square ice array brought to ground state via 325 K thermal demagnetization protocol.

### III. Results and Discussion

Two sets of lithographically defined nano-island ensembles were fabricated in the epitaxial LSMO films: large ensembles of islands arranged in artificial square ice geometries, and smaller ensembles of islands arranged on kagome lattices. The former geometry is comprised of frustrated vertices with four islands, while the latter has frustrated vertices of three islands. We directly compared the magnetic configurations of these square and kagome nano-island ensembles after an alternating magnetic field demagnetization protocol and after a thermal cycling of the sample to the vicinity of its Curie temperature. A comparison of these two methods for an extended LSMO square ice array is shown in Figure 1. After an *ex situ* saturating magnetic field of 100 mT, which gives a white contrast for all horizontal nano-islands, an oscillating magnetic field was brought from 15 mT to zero and a ground-state region of approximately 5 μm diameter is shown in Figure 1(a). The X-PEEM magnetic contrast expected from a ground state configuration is schematically depicted in Figure 1(b), with nano-islands of alternating contrast along the horizontal direction, and neutral contrast along the vertical direction. Due to the slight misorientation between the x-ray incidence direction and the horizontal axis of the square ice array, careful analysis of the vertical nano-island contrast shows a weak alternating contrast which confirms that the center region in Figure 1(a) is in a ground state configuration, while the regions above and below the outlined islands are still oriented along the saturating field direction. However, further repetitions of the alternating magnetic field protocol could not drive the entire array into a ground state. On the other hand, thermal cycling of the sample to 325 K at a rate of 5 K/min brought the square ice arrays into a single ground state configuration across the entire 1200-island arrays. Additional square ice ensembles are shown in the supplemental material (Figure S3), with some isolated high-energy vertices which then can be annealed out after repeated thermal cycling.

While it is clear that the thermal demagnetization protocol can successfully bring square ice arrays into extended ground state configurations, small ensembles based on the kagome ice geometry can be used to compare the two demagnetizing protocols in a more quantitative manner as a function of increasing number of frustrated vertices contained in the ensemble. One-ring hexagonal ensembles have no frustrated vertices, as each nano-island only has two nearest neighbors, and the total dipolar energy landscape of the ensemble can be visualized in a straightforward manner due to the small number of islands.[23] For a six-island ensemble, $2^6 = 64$ possible magnetization configurations exist as each island's magnetization can only orient along its long axis, and the total dipolar energy of the ensemble falls into one of eight degenerate energy levels. As the number of islands in the ensemble increases for two-ring ($2^{11}$ configurations) and three-ring ($2^{15}$ configurations) structures, so too does the magnetic frustration as three-island vertices are now introduced into the ensembles. For the nano-magnet ensembles, the lowest energy magnetic configurations involve vortex or flux closure patterns which have degeneracy (clockwise or counterclockwise orientation per ring), and higher energy configurations involve selected islands reversed when compared to this lowest energy state.[23]

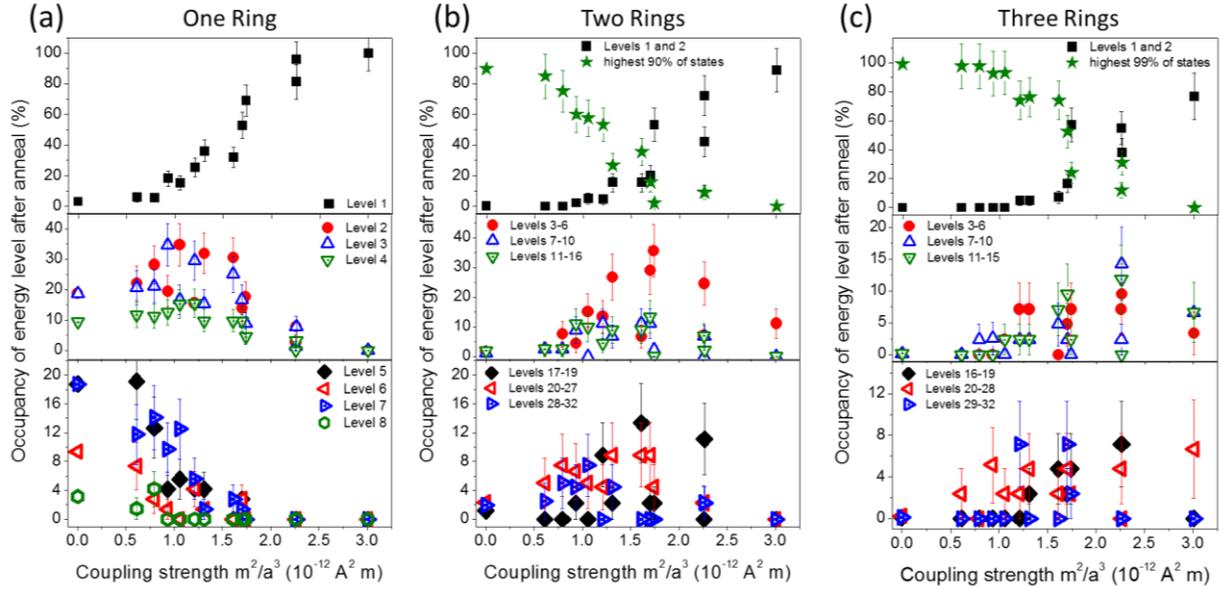

**Figure 2** (double column)– Comparison of the occupancy of low energy magnetization configurations for (a) one-ring, (b) two-ring, and (c) three-ring nano-island ensembles after a 350 K anneal as a function of dipolar coupling strength. The points at zero coupling strength are calculated from the occupancy of energy levels based on the degeneracy of possible random magnetic configurations.

Figure 2 illustrates the occupancy of low-energy states obtained for one-ring, two-ring, and three-ring nano-island ensembles after heating to 350 K (i.e. above their Curie temperature) and cooling back to the measurement temperature of 110 K at a rate of approximately 5 K/min. After the anneal, X-PEEM XMCD images were acquired for twelve sets of one hundred one-, two-, and three-ring ensembles with room-temperature coupling strength $m^2/a^3$ ranging between $6x10^{-13} A\ m^2$ and $3x10^{-12} A\ m^2$. From the X-PEEM images, the dipolar energy of each ensemble was calculated based on the magnetization orientation of all islands in the ensemble. The ensemble configurations were categorized in order of increasing dipolar energy. The energy landscapes are shown in the supplemental material (Figures S4-S6), and they have close lying bands of levels separated by energy gaps. The data has been binned into groups of energy levels for the two-ring and three-ring ensembles due to the large number of possible states compared to the number of ensembles measured. For comparison, data for the one-ring ensembles before the annealing protocol is shown in supplemental material (Figure S7). Error bars are calculated from the square root of the number of observations. In the limit of no coupling between adjacent islands, *e.g.* with $\frac{m^2}{a^3} \sim 6x10^{-13}$ A m², the degeneracy of the magnetization configurations determines the likelihood of an ensemble being in a specific configuration. On the other hand, with strong coupling between neighboring islands, *e.g.* $\frac{m^2}{a^3} \sim 3x10^{-12}$ A m², the ensembles are expected to fall into flux-closure magnetization configurations with a high probability.

Annealing above the Curie temperature allows us to probe the effect that these frustrated vertices have on the probability of the ensembles being in a low energy state. In the regime of high coupling strength, the one-ring nano-island ensembles can effectively be brought into the lowest energy configurations with nearly 100 % probability, but the lowest energy occupancy is noticeably lower in the two-ring (90 %) and three-ring (80%) ensembles. These occupancy levels can be tuned between a nearly random magnetization configuration frequency distribution and having all nano-magnet ensembles fall into the lowest energy level for the one ring ensembles through the choice of lattice spacings between 500-600 nm and island widths of 100-175 nm. As LSMO is cooled through its Curie temperature, the low saturation magnetization results in a weak dipolar interaction between neighboring islands for all ensembles. Assuming that thermally driven transitions between energy levels follow a Boltzmann-like distribution, there is a low energy cost of reversing the magnetization of a single island in the limit of weak dipolar interaction. Such thermally active island reversals can lead to nano-magnet ensembles jumping from a ground state configuration into a higher energy configuration and vice versa. Thus, these thermally active ferromagnetic ensembles can navigate large regions of their energy landscape in a similar manner to thermally active permalloy nano-island ensembles.[11] The above-Curie temperature thermal cycling protocol and 5 K/min cooling allows many ensembles with intermediate room-temperature dipolar coupling strengths of $1 - 2 x 10^{-12}$ A m$^2$ to be frozen into energy configurations slightly higher than the ground state, (*e.g.* Levels 2-4 for the one-ring, Levels 3-6 for the two- and three-ring ensembles). The occupancy of these particular states has a maximum in the intermediate coupling regime, but as the coupling strength decreases, the occupancy converges to the values determined by the degeneracy of the configurations as shown for the data points at the $\frac{m^2}{a^3}$=0 value.

A more carefully designed annealing protocol can increase the overall probability that the ensembles transition to low energy levels by limiting the area of the energy landscape that the ensembles can traverse. While Figure 2 shows the effect of a single annealing protocol on ensembles of different coupling strength, Figure 3 plots the occupation of energy levels for one- and two-ring ensembles at a room-temperature coupling strength of $1.8 \; x 10^{-12}$ A m$^2$. Data for three-ring ensembles can be found in the supplemental material (Figure S8). The ensembles were imaged after an AC magnetic field demagnetization protocol, and subsequently the sample was heated to 325 K for 5 minutes. At this temperature, the sample was near but below the Curie temperature and the film saturation magnetization is approximately 100 emu/cm$^3$. A significant number of the ensembles fell into low energy states after this anneal, higher than the trend seen in Figure 2. A repeat of this anneal pushed the number of low energy configurations to even greater occupancy. An *ex situ* anneal in air to 473 K, well above the Curie temperature, returned the sample to similar energy level probabilities as the 350 K anneal data presented in Figure 2, illustrating that these nano-island ensembles have a higher percentage of low energy configurations when held near but below their Curie temperature. A similar thermally active region was probed with spatially-averaged magneto-optical Kerr effect for delta doped Fe:Pd, with the thermally active region within 25 K of its Curie temperature of 230 K.[12]

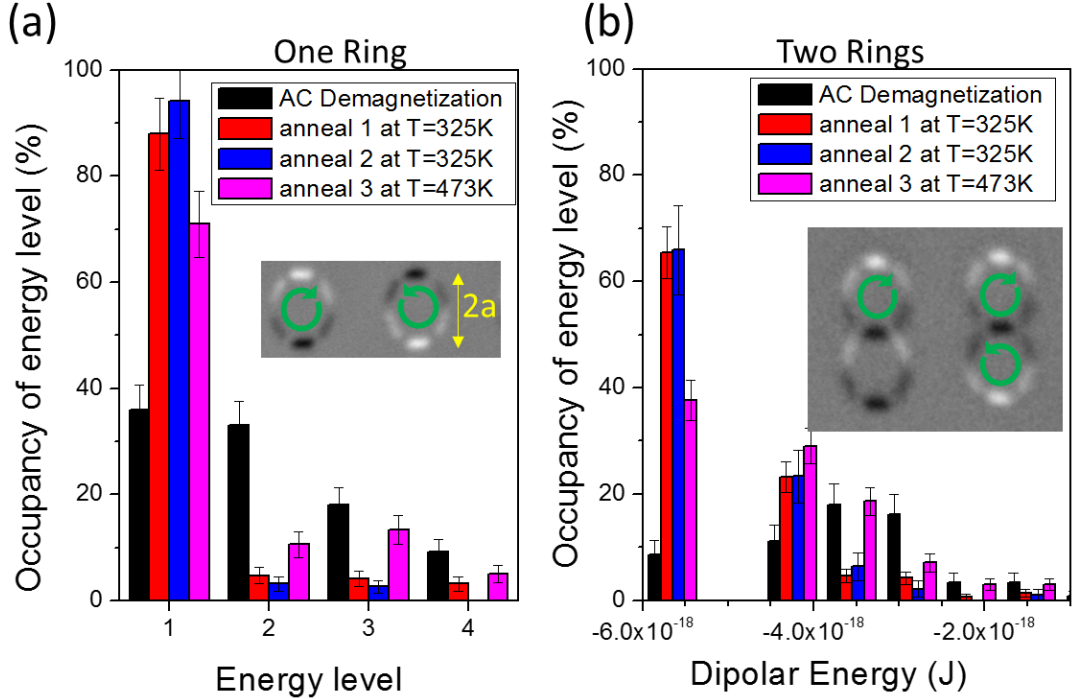

**Figure 3** (single column) – Effect of demagnetizing protocol on occupancy of low energy magnetization configurations for (a) one-ring and (b) two-ring ensembles, with the width of each ring of $2a$=1 µm and a coupling strength of $1.8 \times 10^{-12} A\,m^2$. The configurations are binned by calculating the overall dipolar energy of each ensemble, with X-PEEM images of two low dipolar energy configurations shown as insets and green arrows showing magnetization direction of rings in flux closure arrangements.

To evaluate the ability of an annealing protocol to bring the LSMO nano-island ensembles into a truly thermal state, we took the observed occupancy levels above the ground state level, and examined the relative probability of these excited levels as a function of dipolar coupling strength. Demagnetization protocols for permalloy based artificial spin ice arrays have been parametrized by assuming that the population of isolated high energy vertices in a thermalized array follow a Boltzmann-like distribution.[8,14] The probability of finding an excited vertex in the $i^{th}$ energy level can be described by $P_i(T) = q_i \exp\left(-\frac{E_i}{k_B T}\right)/Z$, with the $i^{th}$ energy level having an energy $E_i$ and multiplicity of configurations $q_i$, and $Z$ as the partition function. We apply a similar description to the ring-based nano-island ensemble energy levels, with the relative occupancy of the excited $i^{th}$ state compared to the lowest energy state can be written as $P_i/P_1 = \frac{q_i}{q_1}\exp\left(-\frac{\Delta E}{k_B T}\right)$. The datasets from Figure 2 were appropriately scaled by the multiplicity and when plotted as a function of the coupling strength (see supplemental material Figure S9), all three types of nano-island ensembles give an effective annealing temperature of 339±2 K. This temperature corresponds to the experimentally observed Curie temperature of the LSMO nanowire within the error of the fitting. Thus, thermal annealing of LSMO nano-island ensembles at 350 K, above the Curie temperature, can bring these nano-island ensembles to Boltzmann-like thermalized distributions of energy levels.

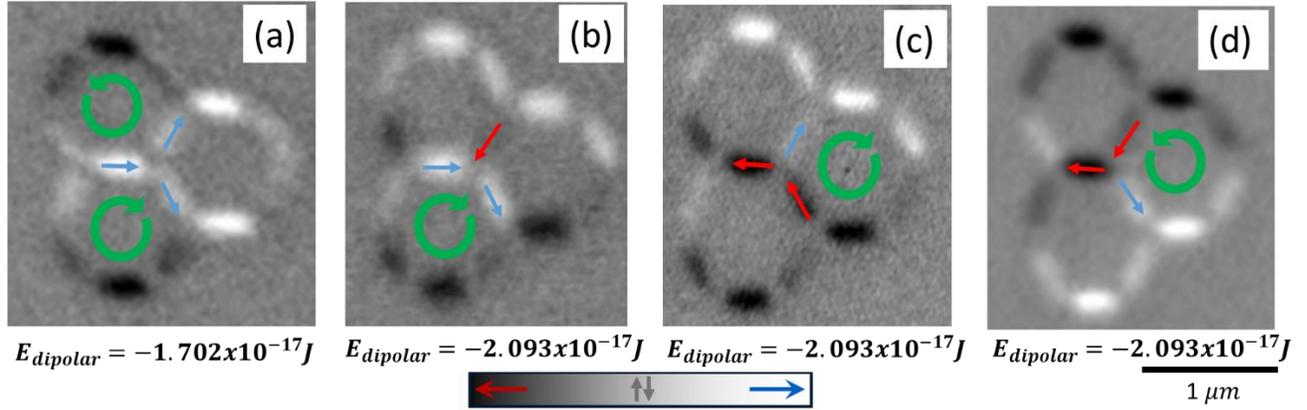

$E_{dipolar} = -1.702 \times 10^{-17} J$    $E_{dipolar} = -2.093 \times 10^{-17} J$    $E_{dipolar} = -2.093 \times 10^{-17} J$    $E_{dipolar} = -2.093 \times 10^{-17} J$

1 µm

**Figure 4** (double column)– X-PEEM images of a three-ring frustrated nano-magnet ensemble (a) after an *in situ* AC magnetic field demagnetization at room temperature, (b) after an *in situ* 5 minute anneal at 325 K, (c) after another *in situ* 5 minute anneal at 325 K, and (d) after a 2 hour *ex situ* anneal in air at 473 K. The green arrows indicate the sense of rotation for island groups that are in a low energy vortex configuration, and the red and blue arrows indicate the direction of magnetization for the center frustrated three-island vertex.

As we can prepare the nano-island ensembles into truly thermalized states through annealing above the Curie temperature, we now directly show real-space measurements of the thermal evolution of a single three-ring ensemble in Figure 4 as it explores low-lying regions of its energy landscape with annealing below the Curie temperature. We followed the same annealing protocol as described for Figure 3. While the perimeter of the three-ring structure falls into a head-to-tail flux closure configuration upon annealing the sample to 325 K (Figure 4 (b)), there still exists a significant degree of freedom in the possible magnetization orientations of the three islands at the center of the ensemble as seen by the repeated annealing at 325 K (Figure 4 (c)). The ensemble has no change in overall dipolar energy between the two 325 K anneals, but a reconfiguration of the center islands can occur while the perimeter islands remain in a fixed configuration. In other words, for annealing steps below the Curie temperature, the ensemble is thermally active, and can escape local minima of the energy landscape in order to explore other configurations with close-lying energy levels. It is only after an anneal well above the Curie temperature that the entire ensemble can traverse its energy landscape in a more global manner such that a majority of the perimeter island orientations can be reversed.

**IV. Conclusions**

In conclusion, complex oxide based artificial spin ice geometries can be used to explore thermally active behavior of magnetization configurations in a similar manner as metallic systems. The chemical stability and epitaxial nature of these materials minimizes experimental issues such as thickness-dependent island-to-island variations and oxidation–induced changes in island magnetization behavior. The magnetization configuration of the nano-island ensembles follow a Boltzmann-like distribution upon annealing the sample through its Curie temperature of 340 K. Additionally, these nano-islands are thermally active at temperatures as low as 325 K and the time-dependent evolution of the magnetization of the ensembles can be imaged in a facile manner, opening up additional avenues for the exploration of thermodynamics of artificial spin ice systems at and above room temperature.

**Acknowledgements**

Part of this work was performed at the Advanced Light Source, Lawrence Berkeley National Laboratory, USA. This research used resources of the Advanced Light Source, which is a DOE Office of Science User Facility under contract no. DE-AC02-05CH11231. Lithography was performed at the Center for Nanophase Materials Sciences, which is a U.S. DOE Office of Science User Facility.

Supplemental material for

Nanostructured complex oxides as a route towards thermal behavior in artificial spin ice systems

R.V. Chopdekar[1], B. Li[1], T.A. Wynn[1], M.S. Lee[1], Y. Jia[1], Z.Q. Liu[2], M.D. Biegalski[2], S.T. Retterer[2], A. T. Young[3], A. Scholl[3], and Y. Takamura[1]

Figure S1 shows resistivity measurement vs. temperature measured on an LSMO nanowire patterned by the same Ar-ion implantation technique used to pattern the nano-island ensembles. The peak in magnetoresistance $MR(H) = \frac{100 \times (\rho(H) - \rho(H=0))}{\rho(H=0)}$ occurs at the Curie temperature ($T_C \sim$ 338 K) due to the double-exchange mechanism responsible for both the electrical and ferromagnetic properties of LSMO.[1]

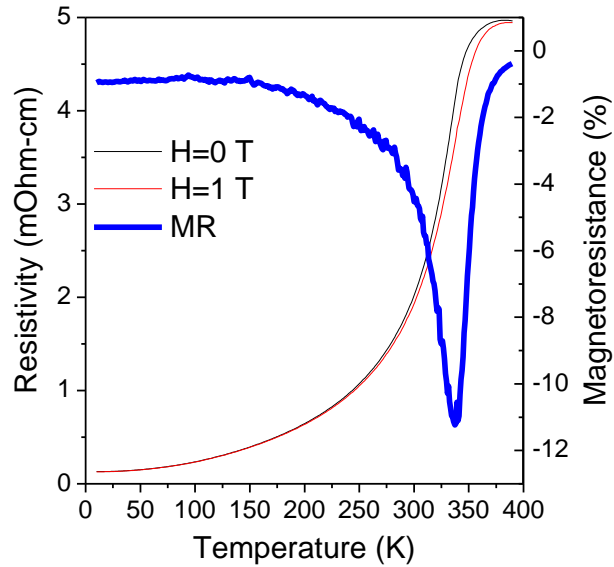

**Figure S1** – Temperature and magnetic-field dependent resistivity of a 500 nm wide LSMO nanowire patterned using the same Ar-ion implant technique as used to pattern the nano-island ensembles.

Figure S2 illustrates the room-temperature magnetic configurations of an as-fabricated one-ring ensemble array measured with magnetic force microscopy (MFM). Contrast in the MFM is determined from the out-of-plane component of the fringing fields emanating from the magnetic nanostructures.[2] The island neighbors that are aligned tail-tail or tip-tip have large fringing fields which give significant dark or bright contrast in the MFM, respectively. Island neighbors arranged in a low-energy tip-tail configuration have small fringing fields at their ends and thus have weak MFM contrast.

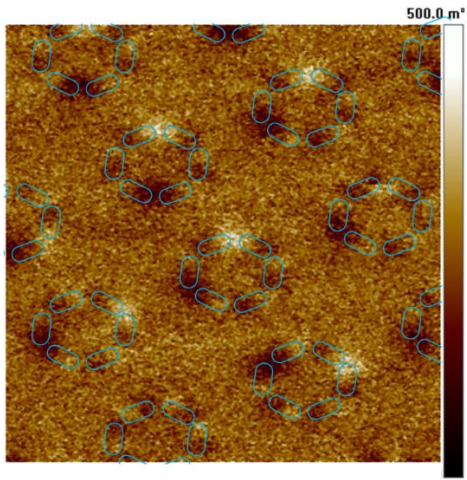

**Figure S2** – Magnetic force microscope image of the one-ring ensembles with coupling strength of $3 \times 10^{-12} A\, m^2$ measured at room temperature. The outlines indicate locations of ferromagnetic LSMO islands, while the rest of the 5 micron field of view is amorphous and paramagnetic.

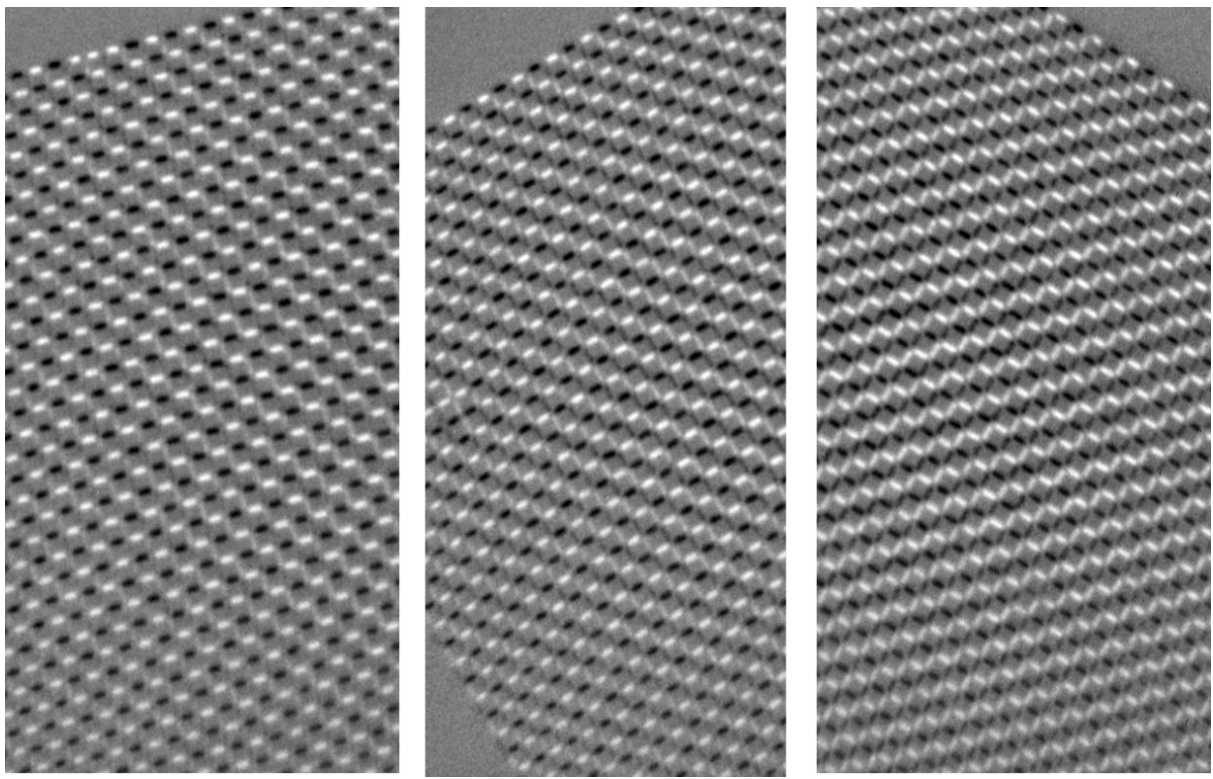

**Figure S3** - Different square ice arrays composed of 470 nm x 175 nm x 40 nm LSMO nano-islands on the same sample as shown in Figure 1 after a 325 K thermal demagnetization, All arrays are in extended ground state configurations with a few isolated excited vertices present.

While there was a distribution of excited states found in the hexagonal-based nano-island ensembles, the square ice arrays (Figure 1 and Figure S3) clearly fell into a low energy configuration with only isolated excited vertices. Upon cooling, if the arrays form local ground state regions that are spatially separated and grow upon cooling, we expect to find the presence of domain walls at the borders of some regions. However, we have not seen any evidence of such domain walls for the large inter-island coupling strength of these arrays, suggesting that high-energy vertices are mobile and a majority can propagate out of the array during the thermal annealing protocol.

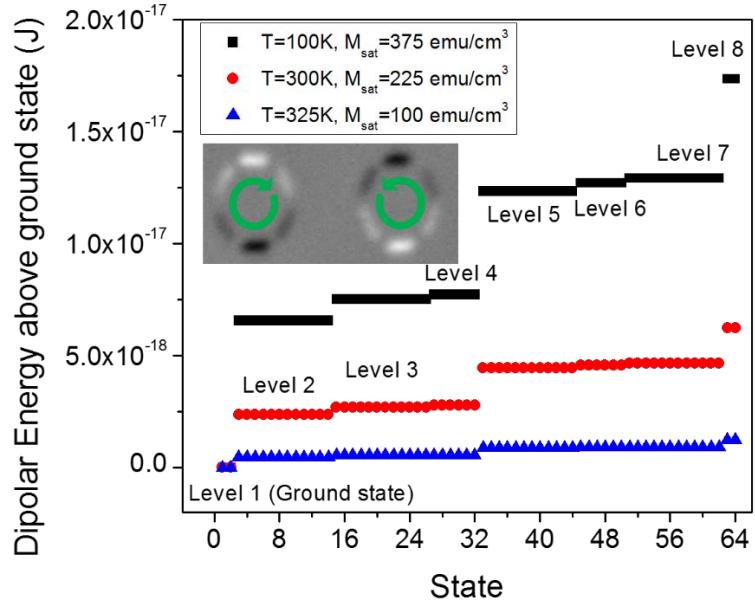

**Figure S4** – Enumerated energy states for the one-ring nano-island ensembles at three measurement temperatures, with energy scale set to an absolute energy of zero for the ground state. Due to the lower saturation magnetization near room temperature, the energy difference between high energy states (Levels 2 and up) and the ground state approaches thermally accessible values. The inset shows X-PEEM example images of the two ground state (Level 1) configurations.

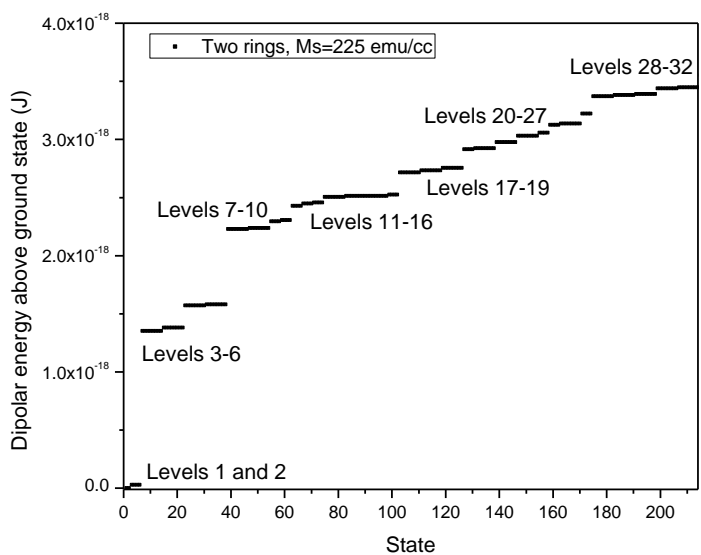

**Figure S5** – Enumerated low-lying energy states (lowest 10% of all possible states) for the two-ring nanoisland ensembles. Specific close-lying states are grouped into energy bands for ease of readability.

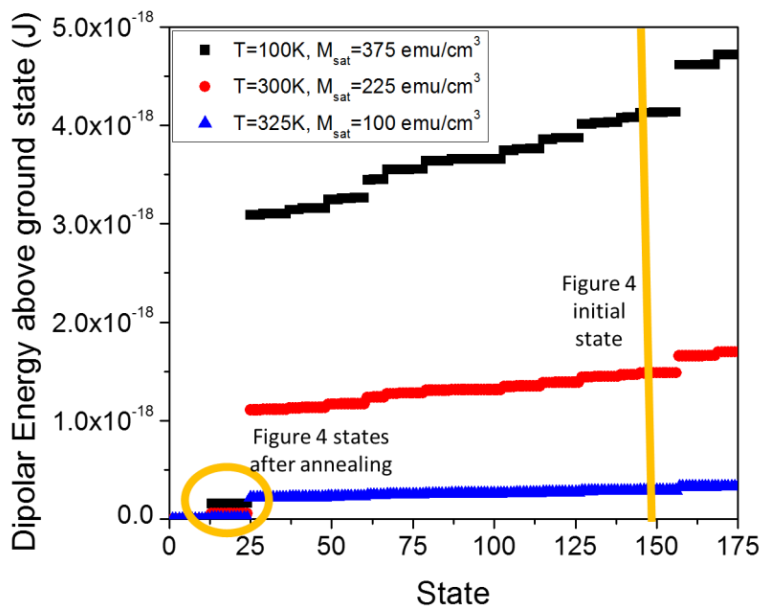

**Figure S6** - Low-lying energy states (lowest 1% of all possible states) for the three-ring nanoisland ensembles, with the configuration of each panel of Figure 4 highlighted.

While each of the 64 possible configurations for the one-ring ensemble can be categorized into one of eight energy levels (see Figures S4, compare to S5 and S6 for the two- and three- ring ensembles), the number of energy levels increases dramatically for the two- and three-ring ensembles.[3] We group the configurations that have similar energy levels into energy bands as the number of ensembles measured in the experiment is much less than the total possible configurations, which results in large errors for counting statistics.

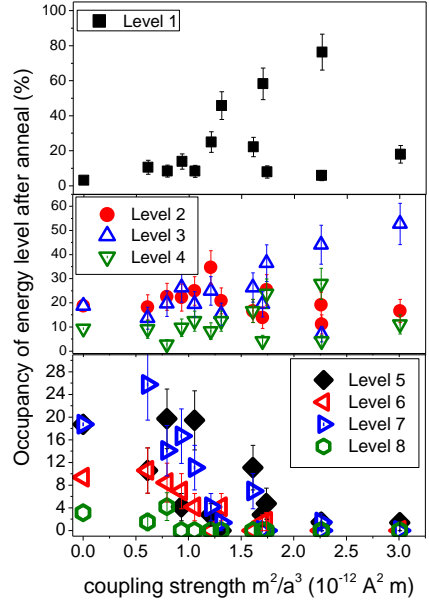

**Figure S7** – Occupancy of energy configurations for one-ring ensembles in the magnetic field demagnetized state before the 350 K anneal. The likelihood of occupancy for the lowest energy configurations (Level 1 in Figure S5) does not follow the monotonic dependence on coupling strength that the post-annealed ensembles demonstrate.

Before the 350 K anneal and acquisition of the data presented in Figure 2, the as-fabricated one-ring ensemble configuration was measured and is plotted in Figure S7. As the sample was held at room temperature for a duration of weeks before the measurement, this suggests that any thermally activated process is extremely slow and the ensembles cannot effectively thermalize at room temperature. The occupancy of the energy levels for the three-ring ensembles is shown in Figure S8, and follows similar trends to the one- and two- ring configurations in Figure 3.

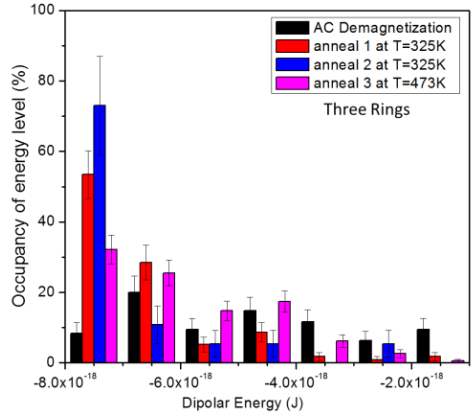

**Figure S8** - Occupancy of low energy magnetization configurations for three-ring ensembles, as a function of type of demagnetizing protocol.

Binning the configurations into energy bands yields a straightforward analysis of the thermalized configurations according to a Boltzmann-like distribution of states.[4, 5] In Figure S9, we plot the relative probability of finding the ensemble in a specific energy band or energy level being occupied as compared to the experimentally found probability of the configuration being in the lowest possible energy level, and plot these relative probabilities as a function of inter-island coupling strength divided by the temperature-dependent magnetization. Plotting the data in this manner allows us to extract the slope as being equal to $M^2(T)/T$, and by using experimentally measured magnetization data for LSMO we can find the effective temperature T of the annealing protocol. The slopes of the curves give an annealing temperature of 339±2 K for all ensembles, and is equal to the Curie temperature within the error of the measurement.

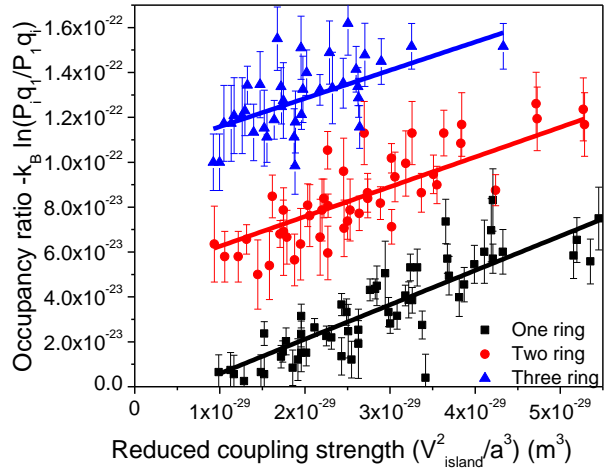

**Figure S9** – Normalized occupancy of excited energy levels as a function of reduced inter-island coupling strength for thermalized ensemble data presented in Figure 2. Assuming a Boltzmann-like probability of energy level occupation, the slope of the trendline should be equal to $M^2(T)/T$, with T as the annealing temperature and M(T) the saturation magnetization at that temperature. Each ensemble dataset is offset vertically for clarity.